\documentclass[12pt]{iopart}
\usepackage{xcolor}
\usepackage{braket}
\usepackage{float}
\usepackage{cite}
\usepackage{graphicx}

\usepackage{iopams}
\begin{document}

\title[Saturated absorption spectra for magnetic field sensing]{Saturated absorption technique used in Potassium microcell for magnetic field sensing}

\author{Armen Sargsyan$^1$, Rodolphe Momier$^{1,2}$, Claude Leroy$^2$ and David Sarkisyan$^1$}
\address{$^1$Institute for Physical Research, NAS of Armenia, Ashtarak-2, 0203 Armenia}
\address{$^2$Laboratoire Interdisciplinaire Carnot de Bourgogne, UMR CNRS 6303, Université Bourgogne Franche-Comté, 21000 Dijon, France}
\ead{rodolphe.momier@u-bourgogne.fr}
\vspace{10pt}
\begin{indented}
\item[]May 2022
\end{indented}

\begin{abstract}
It is demonstrated that the use of a Micrometric Thin $^{39}$K vapor Cell (MTC) and Saturated Absorption spectroscopy (SA) allows to form narrow atomic lines in transmission spectrum without unwanted Cross-Over (CO) resonances. Another important feature is the small characteristic magnetic field value $B_0~=~A_{hf}/\mu_B$ of $^{39}$K, significantly smaller than for Rb and Cs. As a consequence, decoupling of $J$ and $I$ can be observed at relatively low magnetic fields $\sim$300 G, which results in the formation of two groups of four well-spectrally-resolved and equidistantly-positioned atomic transitions having the same amplitude (each group corresponds to a given circular polarization $\sigma^\pm$) which we record using a simple experimental setup with a linearly polarized tunable diode-laser and a longitudinal magnetic field obtained with two permanent magnets. Fabrication of a MTC is much easier than the fabrication of the $^{39}$K nanocells used in our previous works. A simple method to determine the magnitude of a wide range of $B$-fields with a spatial resolution of 30 $\mu$m is presented, which is intrinsically calibrated and does not require a frequency reference.
\end{abstract}

%
\vspace{2pc}
\submitto{\LP}
%
%
%


\section{Introduction}

Atomic Potassium vapors are much less often used than Rb or Cs vapors since even at moderate temperature of about 100 $^\circ$C, the Doppler broadening reaches $\sim$1~GHz. Moreover, the vapor density of K is very small at room temperature (around $5.8~\times~10^{8}$~cm$^{-3}$). Therefore, the hyperfine and Zeeman transition of $^{39}$K transitions turn out to be fully hidden by the Doppler broadening when usual centimeter-long cells are used to study absorption or fluorescence processes. Laser spectroscopy of $^{39}$K was investigated in a rather small number of works. Saturated absorption spectra of $^{39}$K obtained with a 5 cm-long cell were experimentally and theoretically analyzed in \cite{BlochLasPhys6.670}. The magnetically induced dichroism of $^{39}$K $D_2$ line in moderate magnetic fields was studied in \cite{PahwaOptExpr20.17457}. Electromagnetically induced transparency (EIT) in a $^{39}$K vapor was realized in \cite{LampisOptExpr24.15494}, where an EIT-resonance with a linewidth significantly smaller than the natural width ($\sim$ 6 MHz) was formed. Efficient four-wave mixing process in potassium vapors was demonstrated in \cite{ZlatkovicLPL13.015205}. An optical resonance formed on $N$-type level configuration is described in \cite{SargsyanJApplSpectrosc89.12}. 
Splitting of the energy levels of $^{39}$K occurs in an external magnetic field, which leads to the formation of a number of new transitions separated by a frequency interval of $\sim$ 130 MHz. The usage of a Micrometric Thin $^{39}$K vapor cell, as shown below, allows to strongly reduce the Doppler broadening which is an advantage for magnetic field sensing. The advantage of using $^{39}$K is its low characteristic magnetic field value $B_0 = A_{hf}/\mu_B \simeq 165$ G characterizing establishment of the hyperfine Paschen-Back (HPB) regime, where $A_{hf}$ is the magnetic dipole constant for the ground level and $\mu_B$ is the Bohr magneton \cite{OlsenPRA84.063410,ZentileCPC189.162,TiesingaNIST,ArimondoRevModPhys49.31}. Note that $A_{hf}\simeq\Delta/2$, where $\Delta\simeq 462$ MHz is the hyperfine splitting of the ground level $4^2S_{1/2}$ \cite{Tiecke}. $B_0(^{39}\mathrm{K})$ is substantially lower than the analogous value for Rb ($\sim$ 2450 G for $^{87}$Rb, $\sim$ 730 G for $^{85}$Rb) and Cs ($\sim$ 1650 G). Therefore, when $^{39}$K atoms are placed in external magnetic field, important specific features of the behavior of the Zeeman transitions, such as a strong change in their probability and a significant decrease of their number for $B \gg B_0$ (HPB regime) regime \cite{OlsenPRA84.063410,ArimondoRevModPhys49.31,SargsyanEPL110.23001,SargsyanOptLett37.1379,WellerJPB45.215005,SargsyanJETP153.355}) can be easily observed by applying a magnetic field weaker by a factor of 10–15 than for Cs or Rb. In the HPB regime, $J$ and $I$ are decoupled, and only 8 Zeeman transitions remain in the spectrum of the $D_1$ line of $^{39}$K, while the probabilities of the 16 remaining Zeeman transitions tend to zero \cite{SargsyanEPL110.23001}. The splitting of the atomic levels is then described by the magnetic quantum numbers $m_J$ and $m_I$. 
As shown in \cite{Talsky,SargsyanOptLett44.5533,SargsyanPhysLettA434.128043}, taking the second derivative of the absorption spectrum of an atomic vapor allows to obtain complete spectral resolution of the atomic transitions with correct reproduction of both frequency intervals and relative amplitudes. In this work, we will be using this procedure when recording experimental absorption spectra.

\section{Experiment}

\subsection{Micrometric thick $^{39}$K vapor cell}

The windows of the MTC were made of well-polished crystalline sapphire. To minimize birefringence, the windows were cut so that the $C$-axis is perpendicular to their surface. To provide a gap of thickness $L\sim30~\mu$m, thin platinum stripes were placed between the inner surfaces of the windows. In the lower part of the windows, a hole was drilled into which a thin sapphire tube with a diameter of $\sim2$ mm was inserted before gluing. The inner diameter of the tube was 0.8 mm. “Molybdenum” glass glued to the vacuum system was soldered to the sapphire tube. Then, filling the cell with natural potassium is carried out the same way as for glass cells. Details of the construction were presented in \cite{SargsyanOptSpectrosc109.529,SargsyanJETP120.579}. Note that the manufacturing of the nanocell used in \cite{SargsyanOptLett37.1379} is technically a more complicated problem due to the necessity to provide wide regions where the gap thickness is of the order of the wavelength and/or half-wavelength \cite{SargsyanOptLett44.5533}, meanwhile to form the gap in the case of the MTC, it is sufficient to place thin platinum strips (spacers) with a thickness $\sim ~30~\mu$m between the windows. Since manufacturing of this type of MTC also can cause some technical difficulties, the construction was presented in Fig.~2 in paper \cite{SargsyanJETP120.579}. Such cells can be easily manufactured in many laboratories. Other constructions of MTCs were presented for example in \cite{BaluktsianOptLett35.1950,WhittakerJPhysConf635.122006}.

\subsection{Experimental setup}

\begin{figure}[h!]
    \centering
    \includegraphics[scale=0.22]{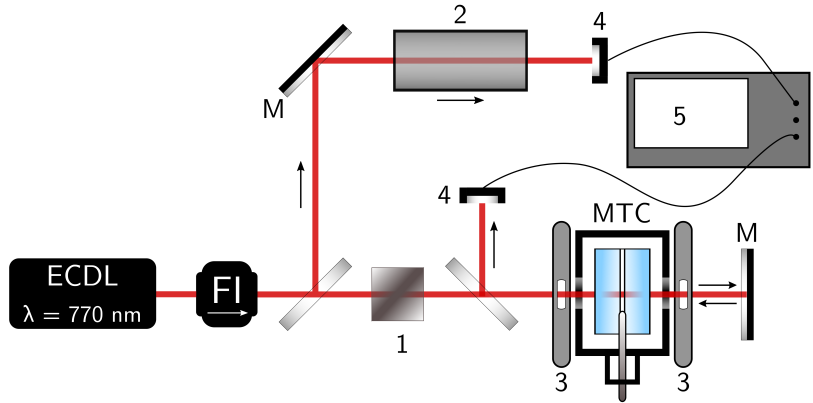}
    \caption{Sketch of the SA experimental setup. ECDL: Extended Cavity Diode-Laser, $\lambda =770$ nm. FI: Faraday Isolator. 1: Glan polarizer. MTC: Micrometric-Thin $^{39}$K Cell. 2: cm-long cell used to form a SA reference spectrum at $B=0$. 3: strong permanent magnets. 4: photodetectors. M: mirrors. 5: digital oscilloscope.}
    
    \label{fig:setup}
\end{figure}

The experimental setup is depicted in Fig.~\ref{fig:setup}. The SA spectrum is recorded using the MTC filled with $^{39}$K, whose thickness is 30 $\mu$m along the laser propagation direction. The MTC was placed into an oven with two holes allowing passage of the laser radiation and was heated to $\sim 130$ $^\circ$C, which provided an atomic density of $N \sim 2\cdot10^{11}$ cm$^{-3}$.

A VitaWave Extended-Cavity Diode Laser (ECDL) with a wavelength of $\lambda = 770$~nm and a spectral linewidth of 1 MHz \cite{VassilievRevSciInstrum77.013102} was used. The MTC was placed between strong permanent magnets (PM) with holes for the passage of laser radiation. The PMs were fixed on nonmagnetic tables. The magnetic field in the MTC was varied by changing the distance between the PMs. The magnetic field was calibrated using a HT201 Teslameter magnetometer based on the Hall effect. A fraction of radiation passing through the MTC was directed precisely backwards using a mirror (M) (in this case, the incident radiation serves as pumping, while the reflected radiation serves as probe radiation) to form the SA spectrum in the MTC. Neutral density filters (not shown in the scheme) were used to determine the optimum pumping and probe radiation powers required for the formation of narrow atomic Velocity Selective Optical Pumping resonances (VSOPs) and reaching their relatively large amplitude at a small spectral width. It has been shown in \cite{SargsyanOptLett39.2270} that so-called Cross-Over resonances (CO) are almost absent in the saturated absorption spectrum of a Rb MTC. This, together with the small spectral width of atomic transitions ($\sim$50 MHz), allows one to use the SA spectrum for the determination of frequencies and probabilities of individual transitions.
To form the frequency reference, a fraction of the laser radiation was directed to a 1.5 cm-long $^{39}$K cell in order to record a SA spectrum at $B=0$.

\subsection{Experimental results and discussion}
	
\begin{figure}[h!]
    \centering
    \includegraphics[scale=0.85]{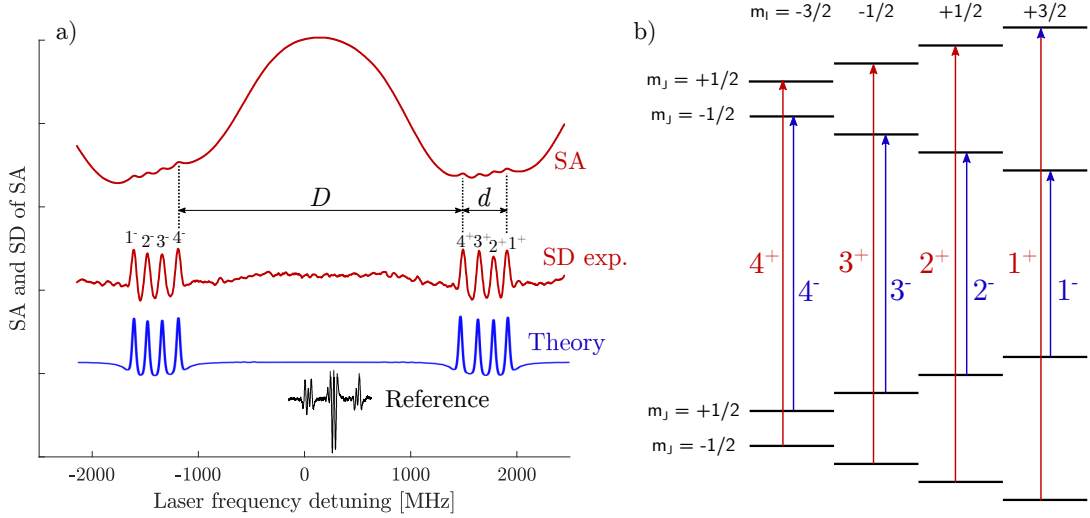}
    \caption{a) Upper curve: experimental SA spectrum of the $D_1$ line of $^{39}$K obtained using a MTC and linearly polarized laser radiation when a longitudinal magnetic field $B = 822$ G is applied. Middle curve: SD of the SA spectrum (inverted for convenience). There are four transitions excited by $\sigma^-$ radiation and four transitions excited by $\sigma^+$ radiation, which are respectively located on the low- and high-frequency wings of the spectrum. Bottom curve: SD of a theoretical absorption spectrum. Black curve: SD of a reference ($B=0$) SA spectrum obtained with a 1.5 cm-long $^{39}$K cell. b) Diagram depicting the $8$ $\sigma^\pm$ transitions present in HPB regime in the uncoupled basis $\ket{m_I,m_J}$. The transitions obey the selection rules $\Delta J = 0$, $\Delta m_I = 0$ and $\Delta m_J = \pm 1$ for $\sigma^\pm$ radiation, respectively shown in red and blue.}
    \label{fig:2}
\end{figure}

In the upper curve of Fig.~\ref{fig:2}a) an experimental SA spectrum of $^{39}$K $D_1$ line obtained using the MTC for linearly polarized laser radiation (consisting of $\sigma^+$ and $\sigma^-$ radiations) and a longitudinal magnetic field $B= 822$ G is presented. The reservoir temperature is 130 $^\circ$C and the laser power is $\sim$ 1mW. The middle curve is the SD of the SA spectrum. Since the condition $B\gg B_0$ required for HPB regime is nearly fulfilled, HPB regime is established. There are four transitions (equidistant in frequency and with the same amplitude) excited by $\sigma^-$ radiation located on the low frequency wing of the spectrum, while another group of four transitions excited by $\sigma^+$ radiation is located on the high-frequency wing of the spectrum. The curve labelled “Theory” is the SD of a theoretical absorption spectrum which correctly shows the frequency position of the 8 transitions and their amplitudes (the theoretical model is described in \cite{TremblayPhysRevA42.2766,DutierJOSAB20.793}). Although the MTC is filled with natural K (93.25\% $^{39}$K, 6.7\% $^{41}$K and 0.01\% $^{40}$K), we neglect the transitions of $^{40}$K and $^{41}$K in the theoretical calculations due do their very small influence on the spectra.
The lower curve is the reference one (SD of SA spectrum) for $B=0$ obtained with the 1.5 cm-long $^{39}$K cell. All the possible $\sigma^+$ transitions labelled $1^+$ to $4^+$ (resp. $\sigma^-$ transitions labelled $1^-$ to $4^-$) in the uncoupled basis $\ket{m_I,m_J}$  are shown in red (resp. blue) in Fig.~\ref{fig:2}b. All transition obey the selection rules $\Delta J = 0$, $\Delta m_I = 0$, and $\Delta m_J = \pm 1$ for the $\sigma^\pm$ transitions.
It is important to note that there are no unwanted CO resonances in the SA and SD spectra, while they are present in the cm-long cells. CO resonances are visible in the reference spectrum. They split under the influence of a magnetic field, leading to undesirable overlapping with the useful atomic resonances. \cite{ZielinskaOptLett37.524}. Their absence is one of the benefits of using a MTC for magnetic field sensing.

\begin{figure}[h!]
    \centering
    \includegraphics[scale=0.8]{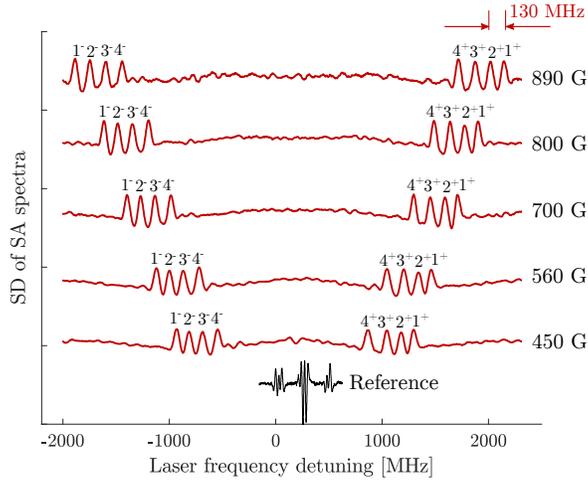}
    \caption{SD of experimental SA spectra obtained using a MTC filled with $^{39}$K probed with linearly polarized laser radiation. The longitudinal magnetic field $B$ is gradually increased from $450$ to $890$ G. There are two pairs of four transitions, one for each circular polarization $\sigma^\pm$. The frequency distance between any two neighboring transitions is $\sim 130$ MHz, as calculated using eq.~(\ref{eq:HPB}). Black curve: SD of a reference ($B=0$) SA spectrum obtained with a 1.5 cm-long $^{39}$K cell.}
    \label{fig:3}
\end{figure}

The SD of experimental SA spectra obtained with the MTC for linearly polarized laser radiation are presented in Fig.~\ref{fig:3}, where the longitudinal magnetic field is gradually increased from 450 to 890 G. As before, the two groups of $\sigma^\pm$ transitions located on the low- and high-frequency wing of the spectra are visible.
The lower curve is the reference spectrum for $B=0$, and the spectra are shifted vertically for clarity.

\begin{figure}[h!]
    \centering
    \includegraphics[scale=0.85]{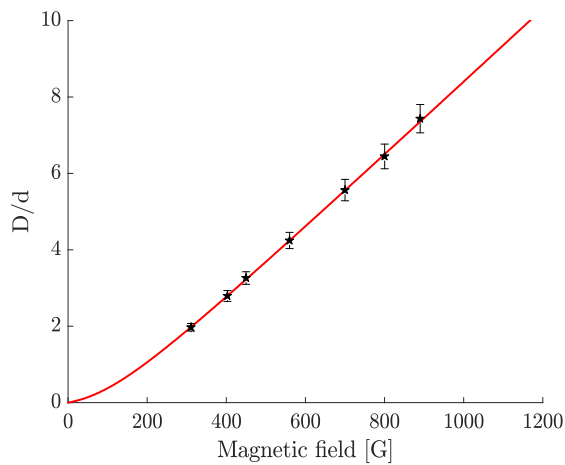}
    \caption{Ratio $D/d$ of the frequency intervals as a function of the magnetic field $B$. Stars with error bars: experimental measurements. Red line: theory. The innacuracy is 5\%.}
    \label{fig:4}
\end{figure}

In a strong magnetic field $B \gg B_0$, the energies of the lower and upper levels can be determined by the expression \cite{SargsyanOptLett37.1379}:

\begin{equation}
    E_{\ket{J,m_J,I,m_I}} = A_{hf}m_Jm_I + \mu_B (g_Jm_J + g_Im_I)B\, , \label{eq:HPB}
\end{equation}
where $A_{hf}$ is the magnetic dipole interaction constant of the lower level $4^2S_{1/2}$ ($\sim$230.86~$h\cdot$MHz) and upper level $4^2P_{1/2}$ ($\sim$27.77~$h\cdot$MHz), and $g_J$ and $g_I$ are the total electronic momentum and nuclear momentum Landé factors, respectively. All the constants related to $^{39}$K are presented in Table B7 in paper \cite{ZentileCPC189.162}. The frequency spacing between any two neighboring transitions estimated with eq.~(\ref{eq:HPB}) is $0.5A_{hf}(4S_{1/2})~+~0.5A_{hf}(4P_{1/2}) ~=~129.3$ MHz. Note that the frequency distance $\sim$130~MHz between two neighboring transitions agrees well with the experiment shown in Fig.~\ref{fig:3}. The frequency slope of both groups of transitions is $s=\mp4\mu_B/3\approx\pm1.86$MHz/G for $\sigma^\pm$ radiation respectively \cite{MomierJQSRT272.107780}. The slope $s$ is an asymptotic value to be reached at $B \gg  B_0$.
Measuring the frequency distance $D$ between transitions labeled $4^+$ and $4^-$ and dividing it by the frequency distance $d$ between transitions $4^+$ and $1^+$ (as depicted in Fig.~\ref{fig:2}), it is possible to determine the magnitude of the $B$-field. The ratio $D/d$ (experiment and theory) as a function of the magnetic field is presented in Fig.~\ref{fig:4}. It should be noted that in this case a spatial resolution of 30 $\mu$m could be fulfilled which is important in the case when a non-homogeneous magnetic field is applied. The ratio $D/d$ was computed by averaging each time over five spectra. More precisely, the spectrum recorded on the oscilloscope is quickly transferred to a computer. A program written for the purpose of this experiment then finds the significant maxima, determines the distances $D$ and $d$ and calculates the ratio $D/d$. The calculation is averaged over five spectra to improve the accuracy. Further, $D$, $d$ and $D/d$ are compared with theoretical calculations to estimate the value of $B$. The inaccuracy in the determination of $D/d$ (and therefore $B$) is caused primarily by the nonlinear scanning of the ECDL lasers over a wide frequency range \cite{VassilievRevSciInstrum77.013102}. The usage of a narrow-band Distributed Feedback (DFB) diode laser as described in \cite{KrastevaPhysScr95.015404}, which has a linear frequency scanning range of $\sim$ 40 GHz, would allow one to use this method of $B$-field measurement in the range 0.1-10 kG.

\section{Conclusion}

The work demonstrates the prospects of using Micrometric Thin $^{39}$K vapor cells to study the peculiarities of the behavior of atomic levels in magnetic fields that are lower by a factor of 10–15 than what is needed to observe the same peculiarities for the Cs or Rb atoms. This is due to the high spectral resolution provided by the Saturation Absorption technique, the absence of unwanted CO resonances (which are present in cm-long cells) and a small magnetic field characteristic value $B_0$. The breakdown of the coupling between $J$ and $I$ can be observed at relatively small magnetic fields $\sim300$ G, which causes the formation of two groups of four equidistantly positioned atomic transitions with the same amplitude that are excited by $\sigma^+$ and $\sigma^-$ radiations. Note that we use a simple scheme involving only one laser.
The paper mentions that the micrometric thick $^{39}$K vapor cell is much easier to manufacture than the $^{39}$K nanocells used in the works \cite{SargsyanEPL110.23001,SargsyanOptLett37.1379} and can be made by a good glass blower following the recommendations given in the paper \cite{SargsyanJETP120.579}. 
A simple method to determine the magnitude of the $B$-field in a wide range with a spatial resolution of 30 $\mu$m is presented and can be carried out without using a frequency reference. Many types of magnetometers are described in the reviews \cite{KitchingApplPhysRev5.031302,FuAVS2.044702}. Optical magnetometers based on nitrogen-vacancy (NV) centers in diamond are extremely sensitive but are quite energy-hungry due to the need of heavy temperature stabilisation and of a microwave coupling field. Moreover, their operational range is limited by the ground-state level anti crossing occurring at $B\sim 1024$ G \cite{WickenbrockApplPhysLett109.053505}. Meanwhile, our atomic magnetometer scheme based on a $^{39}$K microcell is simple and has the advantage of being immune to electric perturbations and thermal drift. Note that the use of the above-mentioned HT201 magnetometer having a sensor area of a few mm$^2$ will lead to large inaccuracies in the determination of magnetic fields with a large gradient. The advantage of our experimental setup compared to a regular Hall gauge magnetometer is the small size of the gauge which is given by the size of the vapor cell. Such type of magnetometer with high spatial resolution could be useful in various fields such as nuclear medicine, NMR and nuclear tomography.
\section*{References}
\bibliographystyle{iopart-num}
\bibliography{biblio.bib}
\end{document}